\documentclass[12pt,epsf]{article}
\usepackage[pctex32]{graphics}
\makeatletter

\@addtoreset{equation}{section}
\newcommand{\be}{\begin{eqnarray}}
\newcommand{\ee}{\end{eqnarray}}
\newcommand{\ba}{\begin{array}}
\newcommand{\ea}{\end{array}}
\newcommand{\nn}{\nonumber}

\makeatletter \@addtoreset{equation}{section} \makeatother

\begin{document}
\vspace{1cm}
\begin{center}
~\\~\\~\\
{\bf  \LARGE Lagrangian of Self-dual Gauge Fields in Various  Formulations}
\vspace{1cm}

                      Wung-Hong Huang\\
                       Department of Physics\\
                       National Cheng Kung University\\
                       Tainan, Taiwan\\

\end{center}
\vspace{1cm}
\begin{center}{\bf  \Large ABSTRACT } \end{center}
The Lagrangian of self-dual gauge theory in various formulations are reviewed.   From these results we see a simple rule and use it to present some new non-covariant  Lagrangian based on the decomposition of spacetime into $D=D_1+D_2+D_3$. Our prescription could be easily extended to more complex decomposition of spacetime and some more examples are presented therefore.  The self-dual property of the new Lagrangian is proved in detail.  We also show that the new non-covariant actions give field equations with 6d Lorentz invariance. 
 
\vspace{2cm}
\begin{flushleft}
*E-mail:  whhwung@mail.ncku.edu.tw\\
\end{flushleft}
%%%%%%%%%%%%%%%%%%%%%%%
\newpage
\section{Introduction}
Chiral p-forms, i.e. antisymmetric boson fields with self-dual (p+1)-form field strengths play a central role in supergravity and in string theory, such as D = 6 and type IIB D = 10 supergravity, heterotic strings [1] and M-theory five-branes  [2].  In particular, they contribute to the ``miraculous'' cancelation of the gravitational anomaly in type-IIB supergravity or superstring theory.  The first calculation of the gravitational anomaly for chiral p-forms was performed in [3] without using a Lagrangian but just guessing suitable Feynman rules that
incorporate the chirality condition. 

It is well known that there is a problem in Lagrangian description of chiral bosons, since manifest duality and spacetime covariance do not like to live in harmony with each other in one action, as first seen by Marcus and Schwarz [4].   Historically, the non-manifestly spacetime covariant action for self-dual  0-form was proposed by Floreanini and Jackiw [5], which is then generalized to p-form by Henneaux and Teitelboim [6].  In general the field strength of chiral p-form $A_{1\cdot\cdot\cdot p}$ is split into electric density $ {\cal E}^{i_1\cdot\cdot\cdot i_{p+1}}$ and magnetic density $ {\cal B}^{i_1\cdot\cdot\cdot i_{p+1}}$:
\be  {\cal E}_{i_1\cdot\cdot\cdot i_{p+1}} &\equiv& F_{i_1\cdot\cdot\cdot i_{p+1}}\equiv \partial_{[{i_1}}A_{i_2\cdot\cdot\cdot i_{p+1}]}\\
 {\cal B}^{i_1\cdot\cdot\cdot i_{p+1}} &\equiv &{1\over (p+1)!}\epsilon^{i_1\cdot\cdot\cdot i_{2p+2}}  F_{i_{p+2}\cdot\cdot\cdot i_{2p+2}}\equiv \tilde F^{i_1\cdot\cdot\cdot i_{p+1}} 
\ee
in which $\tilde F$ is the dual form of $F$.  The Lagrangian is described by
\be  L= {1\over p!} {\vec{\cal B}}\cdot ({\vec{\cal E}}-{\vec{\cal B}})={1\over p!}  \tilde F_{i_1\cdot\cdot\cdot i_{p+1}} ( F^{i_1\cdot\cdot\cdot i_{p+1}} - \tilde F^{i_1\cdot\cdot\cdot i_{p+1}} )
\ee
Note that in order for self-dual fields to exist, i.e. $\tilde F=F$, the field strength $F$ and dual field  strength $\tilde F$ should have the same number of component.  As the double dual on field strength shall give the original field strength the spacetime dimension have to be 2 modulo 4.  Above actions, however, lead to second class constraints and complicates the quantization procedure.  

Siegel in [7]  proposed a manifestly spacetime covariant action of chiral p-form models by squaring the second-class constraints and introducing Lagrange multipliers $\lambda_{ab}$ into the action.  The Lagrangian of chiral 2 form is described by 
\be L_{Siegel}=-{1\over 12} F_{abc}F^{abc}+{1\over4}\lambda_{ab}{\cal F}^{acd}{\cal F}^b_{~cd}
\ee
in which we define
\be {\cal F}\equiv F-\tilde F
\ee
It is easy to see that the field equation $0={\delta S\over \delta \lambda_{ab}}$ implies ${\cal F}=0$ and we get the self-dual property.  Using this property  the other field equation $0={\delta S\over \delta A_{ab}}$ is automatically satisfied.  Siegel action, however, does  not have enough local symmetry to completely gauge the Lagrange multipliers away and suffers from anomaly of gauge symmetry.  

Note that, the self-dual relation $\tilde F=F$ is a first-order differential equation which defines the dynamics of the chiral boson, contrast to other bosonic fields whose equations of motion are usually second-order differential equations. This lead McClain, Wu and Yu to construct chiral field action in a first order form [8]. In this case, for the Lagrange multiplier itself not to carry propagating degrees of freedom one has to introduce an infinite number of auxiliary fields ``compensating'' the dynamics of each other.  However, this infinite set corresponds to the infinite number of local symmetries which cause problems in choosing the right regularization procedure during the quantization. 

Pasti, Sorokin and Tonin in 1995 constructed a Lorentz covariant formulation of chiral p-forms in D = 2(p+1) dimensions that contains a finite number of auxiliary fields in a non-polynomial way [9].  For example,  6D PST Lagrangian is 
\be L_{PST} = -{1\over 6}F_{abc}F^{abc} +{1\over (\partial_q a\partial^q a)} \partial^ma(x) {\cal F}_{mnl}{\cal F}^{nlr}\partial_ra(x)
\ee
in which $a(x)$ is the auxiliary field.  In the gauge $\partial_r a =\delta_r^1$  the PST formulation reduces to the non-manifestly covariant  formulation [5,6]. On the other hand, Perry and Schwarz [10] had shown that the non-covariant action  (1.3) gives field equations with 6d Lorentz invariance.

Recently, a new non-covariant Lagrangian formulation of a chiral 2-form gauge field in 6D, called as (3+3) decomposition,  was derived in [11] from the Bagger-Lambert-Gustavsson (BLG) model [12]. The covariant formulation  of the associated Lagrangian  is constructed in [13], with the use of a triplet of auxiliary scalar fields. Later, a general non-covariant Lagrangian formulation of  self-dual gauge theories in diverse dimensions was constructed [14].  In this general formulation the (2+4) decomposition of  Lagrangian was found. 

In section 2 we review above formulations of  self-dual 2-form in the decomposition of  $D=D_1+D_2$ and find a simple rule.   In section 3 we use the rule to construct new non-covariant actions of self-dual 2-form gauge theory in the decomposition of  $D=D_1+D_2+D_3$.  We present a detailed proof about the self-dual property in the new Lagrangian. We also show in detail that the new non-covariant action gives field equations with 6d Lorentz invariance.  In section 4 we generalize our prescription to construct a  non-covariant action in the  decomposition of $D=D_1+D_2+D_3+D_4$.  Last section is devoted to a short conclusion.  
%%%%%%%%%%%%%%%%%%%
\section {Lagrangian in Decomposition: $D=D_1+D_2$}
To begin with, let us first define a  useful function $L_{ijk}$ :
\be L_{ijk}&\equiv& \tilde F_{ijk}~\times~(F^{ijk} -\tilde F^{ijk}), ~~without~summation~over~indices~i,~j,~k
\ee
which is useful in the following formulations.

%%%%%%%%%%%%%%%%%%%%
\subsection {D=1+5}
In the (1+5) decomposition the spacetime index $A= (1,\cdot\cdot\cdot,6)$ is decomposed as $A=(1,\dot a)$, with $\dot a=(2,\cdot\cdot\cdot,6)$. Then $L_{ABC}=(L_{1\dot a\dot b}, L_{\dot a\dot b\dot c})$.   In terms of $L_{ABC}$, the Lagrangian is expressed as [14]
\be L_{1+5} = -{1\over4} \sum L_{1\dot a\dot b}=- {1\over4} \tilde F_{1\dot a\dot b}(F^{1\dot a\dot b} -\tilde F^{1\dot a\dot b}),~~has~summation~over~\dot a~\dot b
\ee
In table 1 we show all possible form in $L_{ABC}$ and see that $L_{1+5}$ picks up  only $ L_{1\dot a\dot b}$.  Self-dual property of $L_{1+5}$  had been proved in [14].
\\
\\

{Table 1: Lagrangian in various decompositions: $D=D_1+D_2$.}\\\\
\scalebox{1}{\hspace{0.5cm}\includegraphics{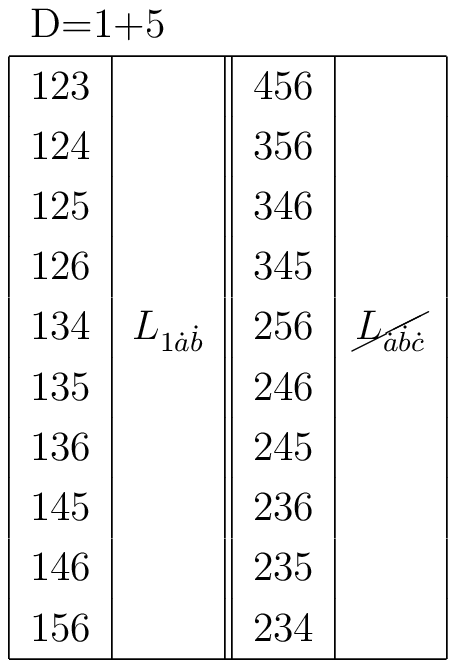}}\scalebox{1}{\hspace{0.5cm}\includegraphics{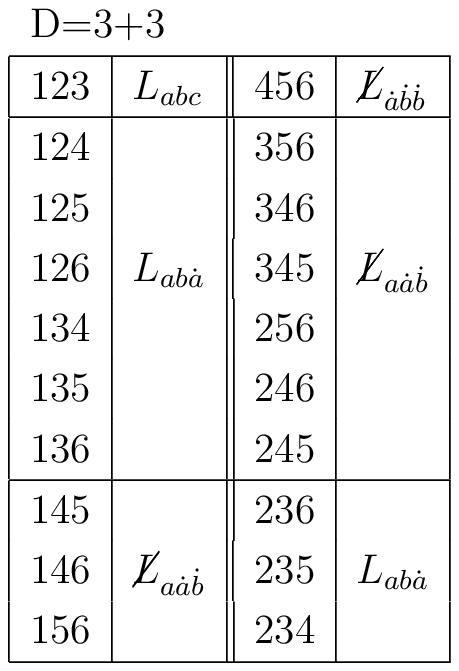}}\scalebox{1}{\hspace{0.5cm}\includegraphics{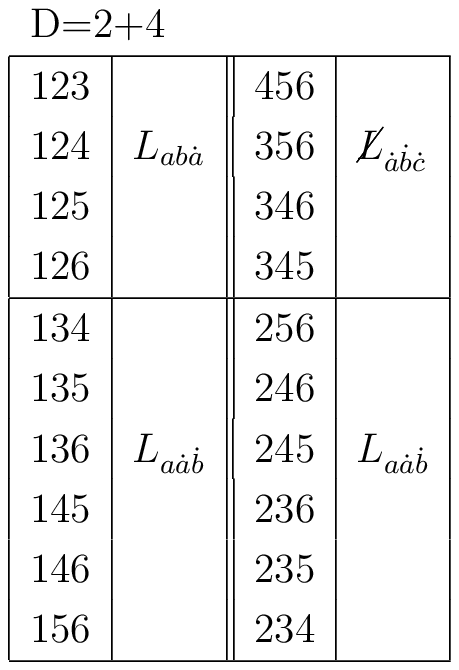}}
%%%%%%%%%%%%%%%%%%%%
\subsection {D=3+3}
In the (3+3) decomposition [14] the spacetime index $A$ is decomposed as $A=(a,\dot a)$, with $a=(1,2,3)$ and $\dot a=(4,5,6)$. Then $L_{ABC}=(L_{abc}, L_{ab\dot a}, L_{a\dot a\dot b}, L_{\dot a\dot b\dot c})$.  Using table 1 it is easy to see that in terms of $L_{ABC}$ the Lagrangian can be expressed as [14]
\be L_{3+3} = -{1\over12} \Big( \sum L_{abc}+3  \sum L_{ab\dot a}\Big)
\ee
Let us make following interesting comments:

1. Why there is the ``3'' factor before $L_{ab\dot a}$ in above equation ? This is because that we have to include three kinds of $L_{ijk}$ : $L_{ab\dot a}$, $L_{a\dot a b}$ and $L_{\dot a ab}$. 

2.  Note that choosing $L_{3+3} \sim  \sum L_{abc}+3  \sum L_{a\dot a\dot b}$ will spoil the gauge symmetry $\delta A_{ab}= \Phi_{ab}$ which is crucial in  proving the self-dual property of the Lagrangian.  A simple rule to have this symmetry is that the choosing Lagrangian $ L_{3+3}$ shall contain all possible  index ``ab" in $L_{ABC}$.  More precisely,  as $L_{ABC}=(L_{abc}, L_{ab\dot a}, L_{a\dot a\dot b}, L_{\dot a\dot b\dot c})$ the all possible term with index ``ab" in $L_{ABC}$ is $L_{abc}, L_{ab\dot a}$.  As both terms have been included in $L_{3+3}$, the Lagrangian thus has the crucial gauge symmetry.  Self-dual property of $L_{3+3}$  had been proved in [13,14].
%%%%%%%%%%%%%%%%%%%%
\subsection {D=2+4}
In the (2+4) decomposition the spacetime index $A$ is decomposed as $A=(a,\dot a)$, with $a=(1,2)$ and $\dot a=(3,\cdot\cdot\cdot,6)$. Then $L_{ABC}=(L_{ab\dot a}, L_{a\dot a\dot b},L_{\dot a\dot b\dot c})$.  From table 1 it is easy to see that in terms of $L_{ABC}$ the Lagrangian can be expressed as [14]
\be L_{2+4} = -{1\over4} \Big( \sum L_{ab\dot a}+{1\over2}  \sum L_{a\dot a\dot b}\Big)
\ee
Self-dual property of $L_{2+4}$  had been proved in [14].  Let us make following interesting comments:

1. Why there is the ${1\over2}$ factor before $L_{a\dot a\dot b}$ in above equation ? This is because that in table 1 $L_{a\dot a\dot b}$ contains both of left-line element and right-line element (for example, it includes $L_{134}$ and $L_{256}$), thus there is double counting. 

2. From table 1 we see that the difference between the Lagrangian in decomposition $D=2+4$ and $D=1+5$ is that we have chosen left-hand (electric) part and right-hand (magnetic) part in $D=2+4$, while in $D=1+5$ we choose only left-hand (electric) part.  In the self-dual theory the electric part is equal to magnetic part.   Thus the Lagrangian choosing electric part is equivalent to that choosing magnetic part. However, in the decomposition into different direct-product of  spacetime one shall choose different part of $L_{ijk}$ to mixing to each other.  This renders the results to be different and we have many kinds of formulation, as shown in the next section.
%%%%%%%%%%%%%%%%%%%
\section {Lagrangian in Decomposition: $D=D_1+D_2+D_3$}
We now consider another decomposition of Lagrangian by: $D=D_1+D_2+D_3$
%%%%%%%%%%%%%%%%%%%%
\subsection {D=1+1+4}
In the (1+1+4) decomposition the spacetime index $A$ is decomposed as $A=(1,2 ,\dot a)$, with $\dot a=(3,4,5,6)$, and  $L_{ABC}=(L_{12\dot a}, L_{\dot a \dot b\dot c}, L_{1\dot a\dot b}, L_{2\dot a\dot b})$.
\\\\
{Table 2: Lagrangian in various decompositions: $D=D_1+D_2+D3$.}\\\\
\scalebox{1}{\includegraphics{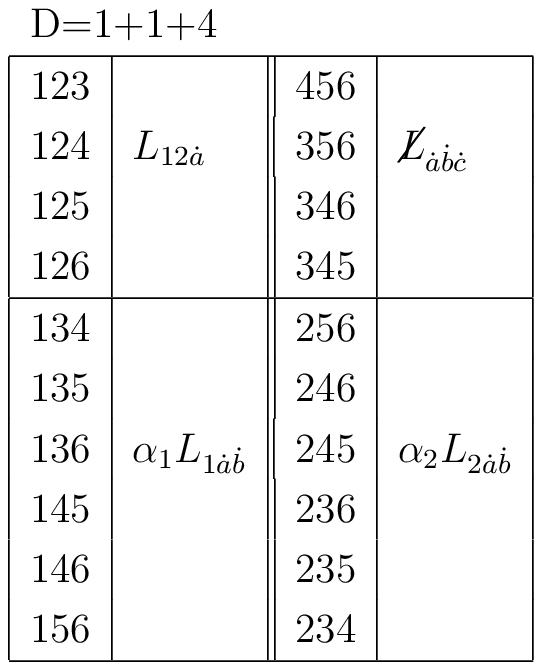}}\scalebox{1}{\hspace{0.5cm}\includegraphics{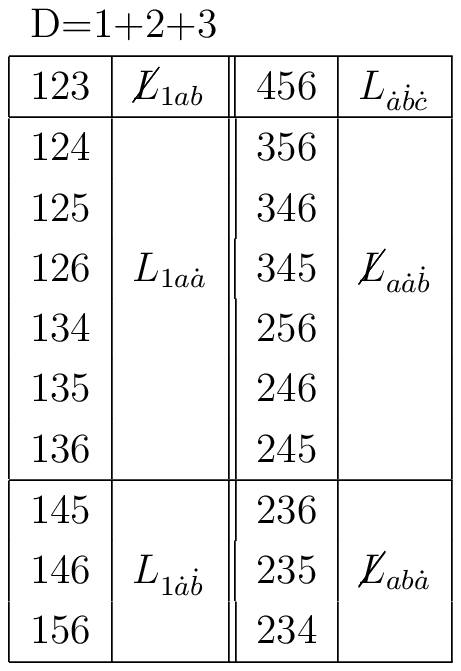}}\scalebox{1}{\hspace{0.5cm}\includegraphics{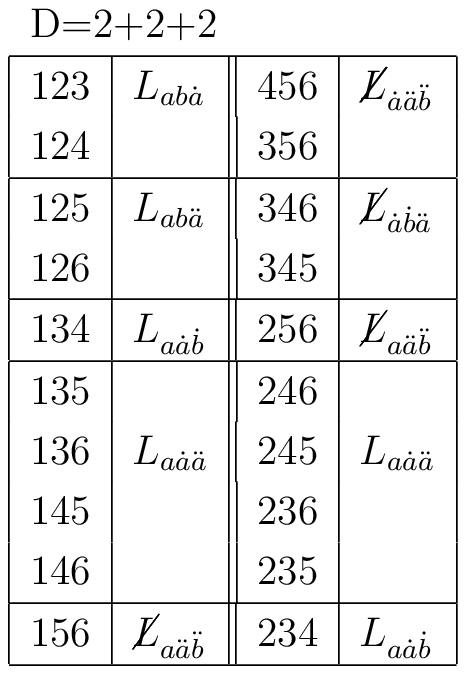}}
\\
\\

From table 2 it is easy to see that, in terms of $L_{ABC}$, the Lagrangian can be expressed as 
\be L_{1+1+4} = 6 \sum L_{12\dot a}+ {3(1-\alpha)\over2}\sum L_{1\dot a\dot b}+{3(1+\alpha)\over2}\sum L_{2\dot a\dot b}
\ee
We neglect overall constant in Lagrangian, which is irrelevant to the following proof.  Note that the case of $\alpha =0$  is just $L_{2+4}$,  the case of $\alpha =-1$  is just $L_{1+5}$, and the case of $\alpha =1$  is just $L_{1+5}$ while  exchanging indices  1 and 2, as can be seen from table 1.

 We now follow the method in [14] to prove the self-dual property of  $L_{1+1+4}$ and follow the method in [10] to prove that the new non-covariant action gives field equations with 6d Lorentz invariance.
%%%%%%%%%%%%%%%%%%%%
\subsubsection {Self-duality in D=1+1+4}
First, we rewrite the Lagrangian as
\be L_{1+1+4} = 6\tilde F_{12\dot a}(F^{12\dot a} -\tilde F^{12\dot a})&+& {3(1-\alpha)\over2}\tilde F_{1\dot a\dot b}(F^{1\dot a\dot b} -\tilde F^{1\dot a\dot b})\nn\\
&+&{3(1+\alpha)\over2}\tilde F_{2\dot a\dot b}(F^{2\dot a\dot b} -\tilde F^{2\dot a\dot b})\nn\\ 
=-\tilde F_{\dot a\dot b\dot c}F^{\dot a\dot b\dot c} +F_{\dot a\dot b\dot c}F^{\dot a\dot b\dot c}&-&{3(1-\alpha)\over2}\tilde F_{2\dot a\dot b}F^{2\dot a\dot b} +{3(1-\alpha)\over2}F_{2\dot a\dot b}F^{2\dot a\dot b}
\nn\\
&+&{3(1+\alpha)\over2}\tilde F_{1\dot a\dot b}F^{1\dot a\dot b} +{3(1+\alpha)\over2}F_{1\dot a\dot b}F^{1\dot a\dot b}
\ee
The variation of the  action $S_{1+1+4}$ gives 
\be {\delta S_{1+1+4}\over \delta A_{12}}=-6 \partial_{\dot a} \tilde F^{12\dot a}=0
\ee
which is identically zero.  This means that terms involved $A_{12}$ only through total derivative terms and we have gauge symmetry
\be   \delta A_{12}=\Phi_{12}
\ee
for arbitrary functions $\Phi_{12}$. The {\bf Gauge symmetry } is crucial to prove the self-duality in following. 

Next, the field equations  
\be 
0={\delta S_{1+1+4}\over \delta A_{1\dot a}}= -6 (\partial_2 \tilde F^{21\dot a}+\partial_{\dot b}\tilde F^{\dot b1a})-6(1+\alpha)\partial_{\dot b} {\cal F}^{\dot b1 \dot a}=-6(1+\alpha)\partial_{\dot b} {\cal F}^{\dot b1 \dot a}\\
0={\delta S_{1+1+4}\over \delta A_{2\dot a}}= -6 (\partial_1 \tilde F^{12\dot a}+\partial_{\dot b}\tilde F^{\dot b2a})-6(1-\alpha)\partial_{\dot b} {\cal F}^{\dot b2 \dot a}=-6(1-\alpha)\partial_{\dot b} {\cal F}^{\dot b2 \dot a}
\ee
has solutions
\be {\cal F}_{1\dot a\dot b}= \epsilon_{12\dot a\dot b\dot c\dot d}\partial^{\dot c}\Phi^{2\dot d}\\
{\cal F}_{2\dot a\dot b}= \epsilon_{12\dot a\dot b\dot c\dot d}\partial^{\dot c}\Psi^{1\dot d}
\ee 
for arbitrary functions $\Phi^{2\dot d}$ and $\Psi^{1\dot d}$.  

We can now follow [14] to find a self-dual relation. First, taking the Hodge-dual
of ${\cal F}_{1\dot a\dot b}$ in above solution and identifying it to the solution ${\cal F}_{2\dot a\dot b}$ in above equation we find that
\be \partial_{\dot a}\Phi_{2\dot b} =  \epsilon_{12\dot a\dot b\dot c\dot d}\partial^{\dot c}\Psi^{1\dot d}
\ee
Acting $\partial^{\dot a}$ on both sides gives 
\be \partial^{\dot a}\partial_{\dot a}\Phi^{2\dot b}=0
\ee 
Following [14],  imposing the boundary condition that the regular field $\Phi^{2\dot b}$ be vanished at infinities will lead to the unique solution $\Phi^{2\dot b}=0$ and we arrive at the self-duality conditions
\be {\cal F}_{1\dot a\dot b}=0
\ee
Taking the Hodge-dual of ${\cal F}_{1\dot a\dot b}$ we also obtain
\be {\cal F}_{2\dot a\dot b}=0
\ee
Using above result the another field equation becomes
\be 0={\delta S_{1+1+4}\over \delta A_{\dot a\dot b}}&=&-3 (\partial_2 \tilde F^{2\dot a\dot b}+\partial_{\dot c}\tilde F^{\dot c \dot a\dot b}+\partial_1 \tilde F^{1\dot a\dot b})\nn\\
&&-6\partial_{\dot c} {\cal F}^{\dot c \dot a\dot b}-3(1+\alpha) \partial_1  {\cal F}^{1\dot a\dot b}-3(1-\alpha) \partial_2  {\cal F}^{2\dot a\dot b}\nn\\
&=& -6\partial_{\dot c} {\cal F}^{\dot c \dot a\dot b}
\ee
which has solution  
\be {\cal F}_{\dot a\dot b\dot c} = \epsilon_{12\dot a\dot b\dot c\dot d}\partial^{\dot d}\Phi^{12}
\ee
We can now use the gauge symmetry of $\delta A_{12}=\Phi_{12}$ to  totally remove $\Phi^{12}$ in ${\cal F}_{\dot a\dot b\dot c}$ and we find a self-dual relation
\be {\cal F}_{\dot a\dot b \dot c}=0
\ee
These complete the proof.
%%%%%%%%%%%%%%%%%%%%
\subsubsection {Lorentz Invariance in D=1+1+4}
As covariant symmetry on 4D coordinates $x_{\dot a}$ is manifest we only need to examine transformations (I) mixing $x_1$ with $x_{\dot a}$, (II) mixing $x_2$ with $x_{\dot a}$ and (II) mixing $x_1$ with $x_2$. 

(I) For the mixing $x_1$ with $x_{\dot a}$ we shall consider the transformation
\be \delta x^{\dot a}&=&\omega^{\dot a 1}~x_1\equiv \Lambda^{\dot a}~x_1,\\
\delta x^{1}&=&\omega^{1\dot a }x_{\dot a}=-\Lambda^{\dot a}~x_{\dot a}=-\Lambda \cdot x
\ee
Define 
\be \Lambda \cdot L\equiv (\Lambda \cdot x)\partial_1 -x_1 (\Lambda \cdot \partial)
\ee
then (detailed in Appendix A)
\be
\delta  F_{12\dot a}&=&(\Lambda \cdot L) F_{12\dot a}+\Lambda^{\dot b} F_{\dot b2\dot a}\\
\delta  F_{\dot a\dot b\dot c}&=&(\Lambda \cdot L) F_{\dot a\dot b\dot c}-\Lambda_{\dot a} F_{1\dot b\dot c}-\Lambda_{\dot b} F_{\dot a 1\dot c}
-\Lambda_{\dot c} F_{\dot a\dot b 1}\\
\delta  F_{1\dot a\dot b}&=&(\Lambda \cdot L) F_{1\dot a\dot b}+\Lambda^{\dot c} F_{\dot c\dot a\dot b}\\
\delta  F_{2\dot a\dot b}&=&(\Lambda \cdot L) F_{2\dot a\dot b}-\Lambda_{\dot a} F_{21\dot b }-\Lambda_{\dot b} F_{2\dot a1}
\ee 
Use above transformation we can find 
\be
\delta  \tilde F_{12\dot a}&=&(\Lambda \cdot L) \tilde F_{12\dot a}+{1\over6}\epsilon_{12\dot a\dot b\dot c\dot d} (\delta_{spin} F^{\dot b\dot c\dot d})\nn\\
&=&(\Lambda \cdot L) \tilde F_{12\dot a}+{1\over6}\epsilon_{12\dot a\dot b\dot c\dot d} [-\Lambda^{\dot b} F^{1\dot c\dot d}-\Lambda^{\dot c} F^{\dot b1\dot d}-\Lambda^{\dot d} F^{\dot b\dot c1}]\nn\\
&=&(\Lambda \cdot L) \tilde F_{12\dot a}+\Lambda^{\dot b} \tilde F_{\dot a\dot b2}\\
\delta  \tilde F_{1\dot a\dot b}&=&(\Lambda \cdot L) \tilde F_{1\dot a\dot b}+{1\over6}\epsilon_{1\dot a\dot b2\dot c\dot d}(\delta_{spin} F^{2\dot c\dot d}\cdot 3)\nn\\
&=&(\Lambda \cdot L) \tilde F_{1\dot a\dot b}+{1\over2}\epsilon_{1\dot a\dot b2\dot c\dot d}[-\Lambda^{\dot c} F^{21\dot d}-\Lambda^{\dot d} F^{\dot 2c1}]\nn\\
&=&(\Lambda \cdot L) \tilde F_{1\dot a\dot b}+\Lambda^{\dot c} \tilde F_{\dot c\dot a\dot b}
\ee 
in which $\delta_{spin} F$ is defined in Appendix A. Therefore 
\be
\delta  (F_{12\dot a}- \tilde F_{12\dot a})&=&(\Lambda \cdot L) (F_{12\dot a}- \tilde F_{12\dot a})+\Lambda^{\dot b}(F_{\dot a\dot b2}- \tilde F_{\dot a\dot b2})\\
\delta (F_{1\dot a\dot b}-\tilde F_{1\dot a\dot b})&=&(\Lambda \cdot L) (F_{1\dot a\dot b}-\tilde F_{1\dot a\dot b})+\Lambda^{\dot c} (F_{\dot c\dot a\dot b}-\tilde F_{\dot c\dot a\dot b})
\ee
which are zero for self-dual theory and the non-covariant action gives field equations with 6d Lorentz transformation mixing $x_1$ with $x_{\dot a}$.

(II) With exchange the index $1\leftrightarrow 2$ above result also shows that the non-covariant action gives field equations with 6d Lorentz transformation mixing $x_2$ with $x_{\dot a}$.

(III) Finally, we consider the mixing $x_1$ with $x_2$.  The transformation is
\be \delta x^{1}&=&\omega^{12}~x_2\equiv \Lambda~x_2,\\
\delta x^{2}&=&\omega^{21}~x_{1}=-\Lambda~x_{1}
\ee
Define 
\be \Lambda \cdot L\equiv (\Lambda x_2)\partial_1 -x_1 (\Lambda  \partial_2)
\ee
then
\be
\delta  F_{12\dot a}&=&(\Lambda \cdot L)F_{12\dot a}\\
\delta  F_{\dot a\dot b\dot c}&=&(\Lambda \cdot L)F_{\dot a\dot b\dot c}\\
\delta  F_{1\dot a\dot b}&=&(\Lambda \cdot L) F_{1\dot a\dot b}-\Lambda F_{\dot a\dot b 2}\\
\delta  F_{2\dot a\dot b}&=&(\Lambda \cdot L) F_{2\dot a\dot b}+\Lambda F_{\dot a\dot b 1}
\ee 
Use above transformation we can calculate the transformations of $\tilde F_{12\tilde a}$ and $\tilde F_{1\dot a \dot b}$. Then we see that
\be
\delta (F_{12\dot a}- \tilde F_{12\dot a})&=&(\Lambda \cdot L)(F_{12\dot a}- \tilde F_{12\dot a})\\
\delta (F_{1\dot a\dot b}-\tilde F_{1\dot a\dot b})&=&(\Lambda \cdot L) (F_{1\dot a\dot b}-\tilde F_{1\dot a\dot b})-\Lambda (F_{\dot a\dot b 2}-\tilde F_{\dot a\dot b 2})
\ee
which are zero for self-dual theory and the non-covariant action gives field equations with 6d Lorentz transformation mixing $x_1$ with $x_2$. 
\\

In summary, we have found the  non-covariant action of self-dual 2-form in decomposition $D=1+1+4$ and have checked that the non-covariant action gives field equations with 6d Lorentz transformation. 
%%%%%%%%%%%%%%%%%%%%
\subsection {D=1+2+3}
In the (1+2+3) decomposition the spacetime index $A$ is decomposed as $A=(1,a ,\dot a)$, with $a=(2,3)$, $\dot a=(4,5,6)$, and  $L_{ABC}=(L_{1ab}$, $L_{1a\dot a}$, $L_{1\dot a\dot b}$, $L_{\dot a\dot b\dot c}$, $L_{a\dot a\dot b}$, $L_{ab\dot a})$.  From table 2 it is easy to see that, in terms of $L_{ABC}$, the Lagrangian can be expressed as 
\be L_{1+2+3} = \sum L_{\dot a\dot b\dot c}+ 6 \sum L_{1a\dot a}+3 \sum L_{1\dot a\dot b}
\ee
Choosing $L_{1ab}+L_{1a\dot a}+L_{1\dot a\dot b}$ is just $L_{1+5}$, and choosing $L_{1ab}+L_{1 a\dot a}+L_{ab\dot a}$ is just $L_{3+3}$, as can be seen from table 1.

We now follow the method in [14] to prove the self-dual property of  $L_{1+2+3}$ and follow the method in [10] to prove that the new non-covariant action gives field equations with 6d Lorentz invariance.
%%%%%%%%%%%%%%%%%%%%
\subsubsection {Self-duality in D=1+2+3}
First, we rewrite the Lagrangian as
\be L_{1+2+3} &=&\tilde F_{\dot a\dot b\dot c}(F^{\dot a\dot b\dot c} -\tilde F^{\dot a\dot b\dot c})+6\tilde F_{1a\dot a}(F^{1a\dot a}-\tilde F^{1a\dot a})+3\tilde F_{1\dot a\dot b}(F^{1\dot a\dot b}-\tilde F^{1\dot a\dot b})\nn\\ 
&=&-3\tilde F_{1ab}F^{1ab} +3F_{1ab}F^{1ab}-3\tilde F_{a\dot a\dot b}F^{a\dot a\dot b}+3F_{a\dot a\dot b}F^{a\dot a\dot b}-3\tilde F_{ab\dot a}F^{ab\dot a}+3F_{ab\dot a}F^{ab\dot a}\nn\\
\ee
The variation of the  action $S_{1+2+3}$ gives
\be {\delta S_{1+2+3}\over \delta A_{1\dot a}}=-6(\partial_a \tilde F^{a1\dot a}+\partial_{\dot b} \tilde F^{\dot b1\dot a}) =0
\ee
which is identically zero. This means that terms involved $A_{1\dot a}$ only through total derivative terms and we have a {\bf gauge symmetry}
\be   \delta A_{1\dot a}=\Phi_{1\dot a}
\ee
for arbitrary functions $\Phi_{1\dot a}$.

 Next, the field equation  
\be 0= {\delta S_{1+2+3}\over \delta A_{\dot a\dot b}}=-3(\partial_{\dot c}\tilde  F^{\dot c\dot a\dot b}+\partial_{1} \tilde F^{1\dot a\dot b}+\partial_{a} \tilde  F^{a\dot a\dot b})-6\partial_{a} {\cal F}^{a\dot a\dot b}=-6\partial_{a} {\cal F}^{a\dot a\dot b}
\ee
has solution
\be {\cal F}^{a\dot a\dot b}= \epsilon^{1ab\dot a\dot b\dot c}\partial_{b}\Phi_{1\dot c}
\ee 
for arbitrary functions $\Phi_{1\dot c}$.   Using the above {\bf gauge symmetry} to completely remove $\Phi_{1\dot c}$ in ${\cal F}_{a \dot a\dot b}$ we obtain  a self-dual relation
\be {\cal F}_{a\dot a\dot b}=0
\ee
To proceed we need to find more gauge symmetry.  First, as term $\Phi_{1\dot c}$ is shown as $\partial_{b}\Phi_{1\dot c}\equiv{\partial\Phi_{1\dot c}\over \partial x^{b} }$  in ${\cal F}^{a\dot a\dot b}$  we have a furthermore symmetry 
\be\delta A_{1\dot a}=W_{1\dot a}(x_1, x_{\dot a})\ee
in which $W_{1\dot a}(x_1, x_{\dot a})$ is an arbitrary function independing on the coordinate $x_{a}$.  In short, after using the gauge symmetry to find a self-dual relation we still have above ``residual gauge symmetries".  Next, as field  $A_{\dot a\dot b}$ only appears as  $\partial_{a}A_{\dot a\dot b}$ in ${\cal F}^{a\dot a\dot b}$, therefore the previous results does not be modified under the variation  
\be\delta A_{\dot a\dot b}= W_{\dot a\dot b}(x_1, x_{\dot a})\ee
in which $W_{\dot a\dot b}(x_1, x_{\dot a})$ is an arbitrary function independing on the coordinate $x_{a}$. We will use the two {\bf residual gauge symmetries} to find other self-duality relations.  Note that these residual gauge symmetries do not spoil any of the self-duality conditions already satisfied.\\

 Now, the field equation  
\be 0= {\delta S_{1+2+3}\over \delta A_{a\dot a}}&=& -6 (\partial_{1}{\tilde F}^{1a \dot a}+\partial_{\dot b}{\tilde F}^{\dot b a \dot a}+\partial_{b}{\tilde F}^{ba \dot a}) + 12(\partial_{\dot b} {\cal F}^{\dot ba\dot a}+\partial_{b} {\cal F}^{ba\dot a})\nn\\
&=&12 (\partial_{\dot b} {\cal F}^{\dot ba\dot a}+\partial_{b} {\cal F}^{ba\dot a})= \partial_{b}  {\cal F}^{ba\dot a}
\ee 
tells us that $ {\cal F}^{ba\dot a}$  is independent of the coordinate$``x_a"$ . 

In the same way, the field equation  
\be 0= {\delta S_{1+2+3}\over \delta A_{1a}}&=& -6 (\partial_{\dot a}{\tilde F}^{\dot a 1a}+\partial_{b}{\tilde F}^{b1 a }) -12\partial_{b} {\cal F}^{b1a}= -12\partial_{b} {\cal F}^{b1a}
\ee 
tells us that $ {\cal F}^{1ab}$  is independent of the coordinate $``x_a"$ . 

Finally, using above properties the field equation  
\be 0= {\delta S_{1+2+3}\over \delta A_{ab}}&=& -3 (\partial_{1}{\tilde F}^{1ab}+\partial_{\dot a}{\tilde F}^{\dot a  ab })+6(\partial_{1} {\cal F}^{1ab}+\partial_{\dot a} {\cal F}^{\dot a ab})\nn\\
&=&6(\partial_{1} {\cal F}^{1ab}+\partial_{\dot a} {\cal F}^{\dot a ab})
\ee
has solution
\be {\cal F}^{1ab}&=& \epsilon^{\dot a\dot b\dot c}\partial _{\dot a}W_{\dot b\dot c}(x_1,x_{\dot a})\\
{\cal F}^{ab\dot a}&=& \epsilon^{\dot a\dot b\dot c}\Big[-\partial _{1}W_{\dot b\dot c}(x_1,x_{\dot a})+ \partial _{\dot b}W_{1\dot c}(x_1,x_{\dot a})\Big]
\ee 
As $W_{\dot b\dot c}$ and  $W_{1\dot c}$ are arbitrary functions independing  on the coordinates $``x_a"$ we can use the residual gauge symmetries to completely remove $W_{\dot b\dot c}$ and  $W_{1\dot c}$.  Thus we obtain the self-dual relations
\be {\cal F}_{1ab}&=&0\\
{\cal F}_{ab\dot a}&=&0
\ee
These complete the proof.
%%%%%%%%%%%%%%%%%%%%
\subsubsection {Lorentz Invariance in D=1+2+3}
As covariant symmetry on 2D coordinates $x_{a}$ and 3D coordinates $x_{\dot a}$ are manifest we only need to examine transformations (I) mixing $x_1$ with $x_{a}$, (II) mixing $x_1$ with $x_{\dot a}$ and (III) mixing $x_a$ with $x_{\dot a}$. 

(I) For the mixing $x_1$ with $x_{\dot a}$ we shall consider the transformation
\be \delta x^{\dot a}&=&\omega^{\dot a 1}~x_1\equiv \Lambda^{\dot a}~x_1,\\
\delta x^{1}&=&\omega^{1\dot a }~x_{\dot a}=-\Lambda^{\dot a}~~x_{\dot a}=-\Lambda \cdot x
\ee
Define 
\be \Lambda \cdot L\equiv (\Lambda \cdot x)\partial_1 -x_1 (\Lambda \cdot \partial)
\ee
then we see that
\be
\delta  (F_{1ab}- \tilde F_{1ab})&=&(\Lambda \cdot L) (F_{1ab}- \tilde F_{1ab})-\Lambda^{\dot a}(F_{ab\dot a}- \tilde F_{ab\dot a})\\
\delta (F_{1a\dot b}-\tilde F_{1a\dot b})&=&(\Lambda \cdot L) (F_{1a\dot b}-\tilde F_{1a\dot b})+\Lambda^{\dot c} (F_{\dot c a\dot b}-\tilde F_{\dot c a\dot b})\\
\delta (F_{1\dot a\dot b}-\tilde F_{1\dot a\dot b})&=&(\Lambda \cdot L) (F_{1\dot a\dot b}-\tilde F_{1\dot a\dot b})+\Lambda^{\dot c} (F_{\dot c \dot a\dot b}-\tilde F_{\dot c \dot a\dot b})
\ee
which are zero for self-dual theory and the non-covariant action gives field equation with 6d Lorentz transformation mixing $x_1$ with $x_{\dot a}$.

%%%%%%%%%%%%%%%
(II) For the mixing $x_1$ with $x_{a}$ we shall consider the transformation
\be \delta x^{a}&=&\omega^{a 1}~x_1\equiv \Lambda^{a}~x_1,\\
\delta x^{1}&=&\omega^{1a }~x_{a}=-\Lambda^{a}~~x_{a}=-\Lambda \cdot x
\ee
Define
\be \Lambda \cdot L\equiv (\Lambda \cdot x)\partial_1 -x_1 (\Lambda \cdot \partial)
\ee
then we see that
\be
\delta  (F_{1ab}- \tilde F_{1ab})&=&(\Lambda \cdot L) (F_{1ab}- \tilde F_{1ab})\\
\delta (F_{1a\dot a}-\tilde F_{1a\dot a})&=&(\Lambda \cdot L) (F_{1a\dot a}-\tilde F_{1a\dot a})+\Lambda^{c} (F_{c a\dot b}-\tilde F_{c a\dot b})\\
\delta (F_{1\dot a\dot b}-\tilde F_{1\dot a\dot b})&=&(\Lambda \cdot L) (F_{1\dot a\dot b}-\tilde F_{1\dot a\dot b})+\Lambda^{c} (F_{c \dot a\dot b}-\tilde F_{c \dot a\dot b})
\ee
which are zero for self-dual theory and the non-covariant action gives field equation with 6d Lorentz transformation mixing $x_1$ with $x_{a}$.
%%%%%%%%%%%%%%

(III) Finally, we consider the mixing $x_a$ with $x_{\dot a}$.  In this case the transformation is 
\be \delta x_{a}&=& \Lambda_a^{~\dot a}~x_{\dot a}\\
\delta x_{\dot a}&=&\Lambda_{\dot a}^{~a}~x_{a}
\ee
Define 
\be \Lambda \cdot L\equiv \Lambda^{a\dot a} (x_a\partial_{\dot a} -x_{\dot a}\partial_{a})
\ee
then we see that
\be
\delta  (F_{1ab}- \tilde F_{1ab})=(\Lambda \cdot L) (F_{1ab}- \tilde F_{1ab})-\Lambda_a^{~\dot a} (F_{1\dot a b}-F_{1\dot a b}) -\Lambda_b^{~\dot b} (F_{1a\dot b}-F_{1a\dot b})\\
\delta (F_{1a\dot b}-F_{1a\dot b})=(\Lambda \cdot L) (F_{1a\dot b}-F_{1a\dot b})-\Lambda_a^{~\dot a} (F_{1\dot a \dot b}-F_{1\dot a \dot b})-\Lambda^b_{~\dot b} (F_{1ab}-F_{1ab})\\
\delta (F_{1\dot a\dot b}-F_{1\dot a\dot b})=(\Lambda \cdot L) (F_{1\dot a\dot b}-F_{1\dot a\dot b})+\Lambda^a_{~\dot a} (F_{1a\dot b}-F_{1a\dot b})+\Lambda^b_{~\dot b} (F_{1\dot a b}-F_{1\dot a b})
\ee
which are zero for self-dual theory and the non-covariant action gives field equation with 6d Lorentz transformation mixing $x_a$ with $x_{\dot a}$.
\\

In summary, we have found the  non-covariant action of self-dual 2-form in decomposition $D=1+2+3$ and have checked  that the non-covariant action gives field equation with 6d Lorentz transformation. 

%%%%%%%%%%%%%%%%%%%
\subsection {D=2+2+2}
In the (2+2+2) decomposition the spacetime index $A$ is decomposed as $A=(a,\dot a,\ddot a)$, with $a=(1,2)$, $\dot a=(3,4)$ and $\ddot a=(5,6)$.  Now, from table 2 we see that  $L_{ABC}=(L_{a\dot a\dot b}$, $L_{\dot a\ddot a\ddot b}$, $L_{ab\ddot a}$, $L_{\dot a\dot b\ddot a}$, $L_{a\dot a\dot b}$, $L_{a\ddot a\ddot b}$, $L_{a\dot a\ddot a})$. Then, in terms of $L_{ABC}$ the Lagrangian can be expressed as 
\be L_{2+2+2} = \sum L_{ab\dot a}+  \sum L_{ab\ddot a}+  \sum L_{a\dot a\dot b}+\sum L_{a\dot a\ddot a}
\ee
We now follow the method in [14] to prove the self-dual property of  $L_{2+2+2}$ and follow the method in [10] to prove that the new non-covariant action gives field equations with 6d Lorentz invariance. 
%%%%%%%%%%%%%%%%%%%%
\subsubsection {Self-duality in D=2+2+2}
 First, we rewrite the Lagrangian as
\be L_{2+2+2} &=& \tilde F_{ab\dot a}(F^{ab\dot a} -\tilde F^{ab\dot a})+\tilde F_{ab\ddot a}(F^{ab\ddot a}-\tilde F^{ab\ddot a})+\tilde F_{a\dot a\dot b}(F^{a\dot a\dot b}-\tilde F^{a\dot a\dot b})\nn\\&&+\tilde F_{a\dot a\ddot a}(F^{a\dot a\ddot a}-\tilde F^{a\dot a\ddot a})\nn\\ 
&=&- \tilde F_{\dot a\ddot a \ddot b}F^{\dot a\ddot a \ddot b}+F_{\dot a\ddot a \ddot b}F^{\dot a\ddot a \ddot b}- \tilde F_{\dot a\dot b\ddot a}F^{\dot a\dot b\ddot a}+F_{\dot a\dot b\ddot a}F^{\dot a\dot b\ddot a} - \tilde F_{ a\ddot a \ddot b}F^{a\ddot a \ddot b}+F_{a\ddot a \ddot b}F^{a\ddot a \ddot b}\nn\\&&- \tilde F_{ a\dot a \ddot b}F^{a\dot a \ddot b}+F_{a\dot a \ddot b}F^{a\dot a \ddot b}
\ee
The variation of the action $S_{2+2+3}$ gives 
\be {\delta S_{2+2+2}\over \delta A_{ab}}=-(\partial_{\dot a}\tilde F^{\dot a ab}+\partial_{\ddot a}\tilde F^{\ddot a ab})=0
\ee
which is identically zero and terms involved $A_{ab}$ only through total derivative terms. Thus, as before, we have a {\bf gauge symmetry} 
\be   \delta A_{ab}=\Phi_{ab}
\ee
for arbitrary functions $\Phi_{ab}$.

 Next, the field equation  
\be 0= {\delta S_{2+2+2}\over \delta A_{\dot a \dot b}}=-(\partial_{a}\tilde F^{a \dot a\dot b}+\partial_{\ddot a}\tilde F^{\ddot a \dot a\dot b})-2\partial_{\ddot a}{\cal F}^{\ddot a \dot a\dot b}=-2\partial_{\ddot a}{\cal F}^{\ddot a \dot a\dot b}
\ee
has solution
\be {\cal F}^{\dot a\dot b\ddot a}= \epsilon^{ab\dot a\dot b\ddot a\ddot b}\partial_{\ddot b}\Phi_{ab}
\ee 
for arbitrary functions $\Phi^{ab}$. As before, we can now use the gauge symmetry of  $A_{ab}$ to reduce ${\cal F}_{\dot a\dot b\ddot a}$ to be zero
\be {\cal F}_{\dot a\dot b\ddot a} =0
\ee
and find the first self-dual relation.  \\

To proceed we need to find more gauge symmetry.  First, as term $\Phi_{ab}$ is shown as $\partial_{\ddot b}\Phi_{ab}$  in ${\cal F}^{\dot a \dot b\ddot a}$  we have a furthermore symmetry 
\be\delta A_{ab}=W_{ab}(x_{a}, x_{\dot a})\ee
in which $W_{ab}(x_a, x_{\dot a})$ is an arbitrary function independing on the coordinate $x_{\ddot a}$. In short, after using the gauge symmetry to find a self-dual relation we still have above  residual gauge symmetries.  Next, as field $A_{\dot a \dot b}$ appears only as $\partial_{\ddot a}A_{\dot a \dot b}$  in ${\cal F}^{\dot a \dot b\ddot a}$ the  above relations do not be modified under the variation  
\be\delta A_{\dot a \dot b}=W_{\dot a \dot b}(x_a, x_{\dot a})\ee
in which $W_{\dot a \dot b}(x_a, x_{\dot a})$ is an arbitrary function independing on the coordinate $x_{a}$. We need  the above two {\bf residual gauge symmetries} to find other self-duality relations in below.   Note that these residual gauge symmetries do not spoil any of the self-duality conditions already satisfied.\\

Now, consider the field equation  
\be 0= {\delta S_{2+2+2}\over \delta A_{a\dot a}}= -2(\partial_{b} \tilde F^{ba\dot a}+\partial_{\dot b} \tilde F^{\dot b a\dot a}+\partial_{\ddot  a} \tilde F^{\ddot a a\dot a})+2\partial_{\ddot  a} {\cal F}^{\ddot a a\dot a}=2\partial_{\ddot  a} {\cal F}^{\ddot a a\dot a}
\ee
which has solution  
\be{\cal F}_{a\dot a\ddot a}=\epsilon_{ab\dot a\dot b\ddot a\ddot b}\partial^{\ddot b}\Phi^{b\dot b}
\ee 
We can now follow [14] to find another self-dual relation.  First, taking the Hodge-dual of both sides in above equation we find that 
\be {\cal F}_{a\dot a\ddot a}=\partial_{\ddot a}\Phi_{a\dot a}
\ee
Identifying above two solutions leads to 
\be \partial_{\ddot a}\Phi_{a\dot a}
= \epsilon_{ab\dot a\dot b\ddot a\ddot b}\partial^{\ddot b}\Phi^{b\dot b}
\ee
Acting $\partial^{\ddot a}$  a on both sides gives
\be \partial^{\ddot a}\partial_{\ddot a}\Phi_{a\dot a}= 0
\ee
Following [14],  imposing the boundary condition that the regular field $\Phi_{a\dot a}$ be vanished at infinities will lead to the unique solution $\Phi_{a\dot a}=0$ and we arrive at the self-duality conditions
\be {\cal F}_{a\dot a\ddot a}=0
\ee
To proceed we need to find one more gauge symmetry.  As term $\Phi_{a\dot a }$ is shown as $\partial_{\ddot a}\Phi_{a\dot a}$  in ${\cal F}^{a\dot a \ddot a}$  we have a furthermore symmetry 
\be\delta A_{a\dot a}=W_{a\dot a}(x_{a}, x_{\dot a})\ee
in which $W_{a\dot a}(x_a, x_{\dot a})$ is an arbitrary function independing on the coordinate $x_{\ddot a}$.  In short, after using the gauge symmetry to find a self-dual relation we still have above  residual gauge symmetries.   We need  the above {\bf residual gauge symmetries} to find other self-duality relations in below. Note that these residual gauge symmetries do not spoil any of the self-duality conditions already satisfied.\\

Use the found self-dual relation the field equation becomes 
\be 0= {\delta S_{2+2+2}\over \delta A_{\dot a\ddot a}}&=& -2(\partial_{\ddot b} \tilde F^{\ddot b\dot a\ddot a}+\partial_{\dot b} \tilde F^{\dot b\dot a\ddot a}+\partial_{a} \tilde F^{a\dot a\ddot a}) + 4\partial_{\ddot b} {\cal F}^{\ddot b\dot b\ddot a}+4\partial_{\dot b} {\cal F}^{\dot b\dot b\ddot a}+4\partial_{a} {\cal F}^{a\dot b\ddot a}\nn\\
&=&4\partial_{\ddot b} {\cal F}^{\ddot b\dot b\ddot a}
\ee
Thus ${\cal F}^{\dot a\ddot a\ddot b}$  is independent of coordinate $x_{\ddot a}$

In a same way, the field equation becomes 
\be 0= {\delta S_{2+2+2}\over \delta A_{a\ddot a}}&=& -2(\partial_{\ddot b} \tilde F^{\ddot b a\ddot a}+\partial_{\dot b} \tilde F^{\dot b a\ddot a}+\partial_{b} \tilde F^{b a\ddot a}) + 4\partial_{\ddot b} {\cal F}^{\ddot b a\ddot a}+2\partial_{\dot a} {\cal F}^{\dot a a\ddot a}\nn\\
&=&4\partial_{\ddot b} {\cal F}^{\ddot b a\ddot a}
\ee
Thus ${\cal F}^{a\ddot a\ddot b}$  is independent of coordinate $x_{\ddot a}$

Use above property of independent of coordinate $x_{\ddot a}$ the final field equation  
\be 0= {\delta S_{2+2+2}\over \delta A_{\ddot a\ddot b}}&=& -(\partial_{\dot a} \tilde F^{\dot a \ddot a\ddot b}+\partial_{a} \tilde F^{a\ddot a\ddot b}) + 2\partial_{\dot a} {\cal F}^{\dot a\ddot a\ddot b}+2\partial_{a} {\cal F}^{a\ddot a\ddot b}\nn\\
&=&2\partial_{\dot a} {\cal F}^{\dot a\ddot a\ddot b}+2\partial_{a} {\cal F}^{a\ddot a\ddot b}
\ee
gives the solution
\be {\cal F}^{a\ddot a\ddot b}&=& \epsilon^{ab\dot a\dot b\ddot a\ddot b}[\partial_b W_{\dot a \dot b}+\partial_{\dot a} W_{b \dot b}]\\
{\cal F}^{\dot a\ddot a\ddot b}&=& \epsilon^{ab\dot a\dot b\ddot a\ddot b}[\partial_a W_{b \dot b}+\partial_{\dot b} W_{ab}]
\ee 
As $W_{\dot a\dot b}$,  $W_{a \dot a}$ and  $W_{ab}$ are arbitrary functions independing on the coordinates $``x_{\ddot a}"$ we can use the ``residual gauge symmetries" to completely remove them.  Thus we obtain the self-dual relations
\be {\cal F}^{a\ddot a\ddot b}&=&0\\
{\cal F}^{\dot a\ddot a\ddot b}&=&0
\ee
These complete the proof.
%%%%%%%%%%%%%%%%%%%%
\subsubsection {Lorentz Invariance in D=2+2+2}
As covariant symmetry on 2D coordinates $x_{a}$, 2D coordinates $x_{\dot a}$ and 2D coordinates $x_{\ddot a}$ are manifest we only need to examine transformations (I) mixing $x_a$ with $x_{\dot a}$, (II) mixing $x_{\dot a}$ with $x_{\ddot a}$ and (III) mixing $x_a$ with $x_{\ddot a}$.

We consider the mixing $x_a$ with $x_{\dot a}$.  In this case the transformation is 
\be \delta x_{a}&=& \Lambda_a^{~\dot a}~x_{\dot a}\\
\delta x_{\dot a}&=&\Lambda_{\dot a}^{~a}~x_{a}
\ee
Define 
\be \Lambda \cdot L\equiv \Lambda^{a\dot a} (x_a\partial_{\dot a} -x_{\dot a}\partial_{a})
\ee
then
\be
\delta  F_{ab\dot a}&=&(\Lambda \cdot L) F_{ab\dot a}-\Lambda_a^{~\dot b} F_{\dot b b\dot a}-\Lambda_b^{~\dot b} F_{a\dot b\dot a}\\
\delta  F_{ab\ddot a}&=&(\Lambda \cdot L) F_{ab\ddot a}-\Lambda_a^{~\dot b} F_{\dot b b\ddot a}-\Lambda_b^{~\dot b} F_{a\dot b\ddot a}\\
\delta  F_{a\dot a\dot b}&=&(\Lambda \cdot L) F_{a\dot a\dot b}+\Lambda^b_{~\dot a} F_{ab\dot b}+\Lambda^b_{~\dot b} F_{a\dot a b}\\
\delta  F_{a\dot a\ddot a}&=&(\Lambda \cdot L) F_{a\dot a\ddot a}-\Lambda_a^{~\dot b} F_{\dot b\dot a\ddot a}-\Lambda^b_{~\dot a} F_{a b\ddot a}\\
\delta  F_{a\ddot a\ddot b}&=&(\Lambda \cdot L) F_{a\ddot a\ddot b}-\Lambda_a^{~\dot a} F_{\dot a \ddot a\ddot b}\\
\delta  F_{\dot a\ddot a\ddot b}&=&(\Lambda \cdot L) F_{\dot a\ddot a\ddot b}+\Lambda^a_{~\dot a} F_{a\ddot a \ddot b}\\
\delta  F_{\dot a\dot b\ddot a}&=&(\Lambda \cdot L) F_{\dot a\dot b\ddot a}+\Lambda^a_{~\dot a} F_{a\dot b\ddot a}+\Lambda^b_{~\dot b} F_{\dot a b\ddot a}
\ee 
Use above transformation we can calculate the transformations of $\tilde F_{ab\dot a}$,  $\tilde F_{ab\ddot a}$, $\tilde F_{a\dot a \dot b}$ and $\tilde F_{a\dot a \ddot a}$. Then we see that
\be
\delta  (F_{ab\dot a}-\tilde F_{ab\dot a})=(\Lambda \cdot L) (F_{ab\dot a}-\tilde F_{ab\dot a})-\Lambda_a^{~\dot b} (F_{\dot b b\dot a}-\tilde F_{\dot b b\dot a})-\Lambda_b^{~\dot b} (F_{a\dot b\dot a}-\tilde F_{a\dot b\dot a})\\
\delta  (F_{ab\ddot a}-\tilde F_{ab\ddot a})=(\Lambda \cdot L)(F_{ab\ddot a}-\tilde F_{ab\ddot a})-\Lambda_a^{~\dot b} (F_{\dot b b\ddot a}-\tilde F_{\dot b b\ddot a}) -\Lambda_b^{~\dot b} (F_{a\dot b\ddot a}-\tilde F_{a\dot b\ddot a})\\
\delta  (F_{a\dot a\dot b}-\tilde F_{a\dot a\dot b})=(\Lambda \cdot L) (F_{a\dot a\dot b}-\tilde F_{a\dot a\dot b})+\Lambda^b_{~\dot a} (F_{ab\dot b}-\tilde F_{ab\dot b})+\Lambda^b_{~\dot b} (F_{a\dot a b}-\tilde F_{a\dot a b})\\
\delta (F_{a\dot a\ddot a}-\tilde F_{a\dot a\ddot a})=(\Lambda \cdot L) (F_{a\dot a\ddot a}-\tilde F_{a\dot a\ddot a})-\Lambda_a^{~\dot b} (F_{\dot b\dot a\ddot a}-\tilde F_{\dot b\dot a\ddot a})-\Lambda^b_{~\dot a} (F_{a b\ddot a}-\tilde F_{a b\ddot a})
\ee
which are zero for self-dual theory and the non-covariant action gives field equations with 6d Lorentz transformation mixing $x_a$ with $x_{\dot a}$. The transformation mixing $x_a$ with $x_{\ddot a}$ and mixing $x_{\dot a}$ with $x_{\ddot a}$ have the similar results. 
\\

In summary, we have found the  non-covariant action of self-dual 2-form in decomposition $D=2+2+2$ and checked  that the non-covariant action gives field equations with 6d Lorentz transformation. 

%%%%%%%%%%%%%%%%%%%%%%
\section {Other Decomposition and Spacetime}
\subsection {Other Decomposition : D=1+1+1+3}
Besides the decomposition in section 2 and section 3 there are many other possible decompositions.  We will in this subsection see that it is possible to found the Lagrangian of self-dual 2 form in decomposition : $D=1+1+1+3$.\\

In the (1+1+1+3) decomposition the spacetime index $A$ is decomposed as $A=(1,2,3,\dot a)$, with $\dot a=(4,5,6)$,and  $L_{ABC}$= ($L_{123}$, $L_{12\dot a}$, $L_{13\dot a}$, $L_{1\dot a\dot b}$, $L_{\dot a \dot b\dot c}$, $L_{2\dot a\dot b}$, $ L_{3\dot a\dot b}$, $L_{23\dot a}$).  From table 3 it is easy to see that, in terms of $L_{ABC}$, the Lagrangian can be expressed as 
\be L_{1+1+1+3} = 6 L_{123}+6 \sum L_{12\dot a}+6 \sum L_{13\dot a}+{3\over2} \sum L_{1\dot a\dot b}+ {6\over2} \sum L_{23\dot a}
\ee
Other choices will becomes the decompositions in section 2 or section 3, as can be seen from table 1 and table 2.  Note that the factor $6$ or $3$ in (4.1) is to count the number of possible permutation and the factor ${1\over2}$ in (4.1) revels that fact we have counted both the left-hand side and right-hand side in table 3.
\\\\
{Table 3: Lagrangian in decompositions: $D=1+1+1+3$.}\\\\
\scalebox{1}{\hspace{5cm}\includegraphics{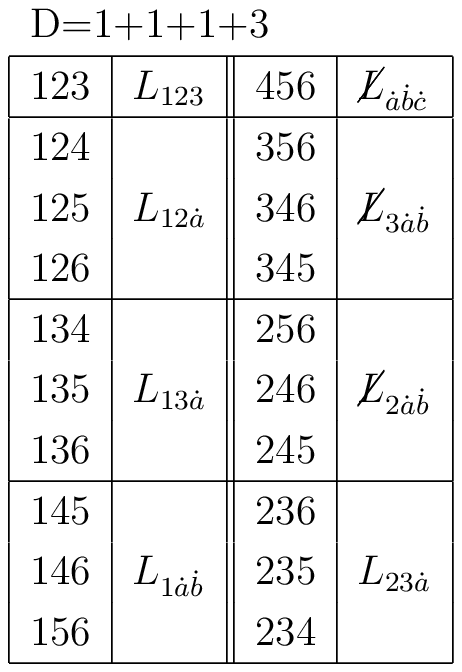}}
\\

We follow the method in [14] to prove the self-dual property of  $L_{1+1+1+3}$ and follow the method in [10] to prove that the new non-covariant action gives field equations with 6d Lorentz invariance.
%%%%%%%%%%%%%%%%%%%%
\subsubsection {Self-duality in D=1+1+1+3}
First, we rewrite the Lagrangian as
\be L_{1+1+1+3} &=& 6\tilde F_{123}(F^{123} -\tilde F^{123})+6\tilde F_{12\dot a}(F^{12\dot a} -\tilde F^{12\dot a})+6\tilde F_{13\dot a}(F^{13\dot a} -\tilde F^{13\dot a})\nn\\
&&~~~~~~~~~+{3\over2}\tilde F_{1\dot a\dot b}(F^{1\dot a\dot b} -\tilde F^{1\dot a\dot b})+{6\over2} \tilde F_{23\dot a}(F^{23\dot a} -\tilde F^{23\dot a})\nn\\ 
 &=& -\tilde F_{\dot a\dot b\dot c}F^{\dot a\dot b\dot c}+ F_{\dot a\dot b\dot c}F^{\dot a\dot b\dot c} -3\tilde  F_{3\dot a\dot b}F^{3\dot a\dot b}+3 F_{3\dot a\dot b}F^{3\dot a\dot b}- 3\tilde F_{2\dot a\dot b}F^{2\dot a\dot b}+ 3F_{2\dot a\dot b}F^{2\dot a\dot b}\nn\\
&& - {3}\tilde F_{23\dot a}F^{23\dot a}+ {3}F_{23\dot a}F^{23\dot a}- {3\over2}\tilde F_{1\dot a\dot b}F^{1\dot a\dot b}+ {3\over2}F_{1\dot a\dot b}F^{1\dot a\dot b}
\ee
The variation of the action $S_{1+1+1+3}$ gives
\be {\delta S_{1+1+1+3}\over \delta A_{12}}&=&-6\partial_{3} \tilde F^{312}-6\partial_{\dot a} \tilde F^{\dot a 12}=0,\\
{\delta S_{1+1+1+3}\over \delta A_{13}}&=&-6\partial_{2} \tilde F^{213}-6\partial_{\dot a} \tilde F^{\dot a 13}=0
\ee
which are identically zero. This means that terms involved $A_{12}$ and $A_{13}$  only through total derivative terms and we have a gauge symmetry
\be   \delta A_{12}=\Phi_{12}\\
 \delta A_{13}=\Phi_{13}
\ee
for arbitrary functions $\Phi_{12}$ and $\Phi_{13}$.

We also have following two field equations  
\be 0= {\delta S_{1+1+1+3}\over \delta A_{1\dot a}}&=&-6(\partial_{2} \tilde F^{21\dot a}+\partial_{3} \tilde F^{31\dot a}+\partial_{\dot b} \tilde F^{\dot b1\dot a})+ 6\partial_{\dot b} {\cal F}^{\dot b1\dot a}=6\partial_{\dot b} {\cal F}^{\dot b1\dot a}\\
 0= {\delta S_{1+1+1+3}\over \delta A_{23}}&=&-6(\partial_{1} \tilde F^{123}+\partial_{\dot a} \tilde F^{\dot a 23})+ 6\partial_{\dot a} {\cal F}^{\dot a 23}=6\partial_{\dot a} {\cal F}^{\dot a 23}
\ee
which imply the solutions 
\be
{\cal F}^{1\dot a\dot b}&=&\epsilon^{123\dot a\dot b\dot c}\partial_{\dot c}\Phi_{23}\\
{\cal F}^{23\dot a}&=&\epsilon^{123\dot a\dot b\dot c}\partial_{\dot b}\Phi_{1\dot c}
\ee
Take the Hodge-dual of  the first solution and compare it with the second solution we find that 
\be  \partial _{\dot a}\Phi_{1\dot b}=\epsilon_{123\dot a\dot b\dot c}\partial ^{\dot c}\Phi_{23}
\ee
After acting $\partial^{\dot a}$ on both sides we find that 
\be \partial^{\dot a}\partial _{\dot a}\Phi_{1\dot b}=0
\ee
As described in [14], imposing the boundary condition that the regular field $\Phi_{1\dot b}$ be vanished at infinities leads to the unique solution $\Phi_{1\dot a}=0$.  Then we arrive at the self-duality conditions
\be {\cal F}_{23\dot a}=0
\ee
In a similar way we can also find another self-duality conditions
\be {\cal F}_{1\dot a\dot b}=0
\ee

Using above result the field equation 
\be 0= {\delta S_{1+1+1+3}\over \delta A_{2\dot a}}&=&-6(\partial_{1} \tilde F^{12\dot a}+\partial_{3} \tilde F^{32\dot a}+\partial_{\dot b} \tilde F^{\dot b2\dot a}) +6\partial_{3} {\cal F}^{32\dot a}+12 \partial_{\dot b} {\cal F}^{\dot b2\dot a}\nn\\
&=& 6\partial_{3} {\cal F}^{32\dot a}+12 \partial_{\dot b} {\cal F}^{\dot b2\dot a}=12 \partial_{\dot b} {\cal F}^{\dot b2\dot a}
\ee 
implies 
\be 
{\cal F}_{2\dot a\dot b}= \epsilon_{123\dot a\dot b\dot c}\partial{\dot c}\Phi_{13}
\ee
Thus, use the gauge symmetry  $\delta A_{13}=\Phi_{13}$ we can completely remove the function $\Phi_{13}$ and find the self-dual relation.
\be {\cal F}_{2\dot a\dot b}=0
\ee 
In the same way, the field equation 
\be 0= {\delta S_{1+1+1+3}\over \delta A_{3\dot a}}&=&-6(\partial_{1} \tilde F^{13\dot a}+\partial_{2} \tilde F^{23\dot a}+\partial_{2} \tilde F^{23\dot a}) +6\partial_{2} {\cal F}^{23\dot a}+12 \partial_{\dot b} {\cal F}^{\dot b3\dot a}\nn\\
&=& 6\partial_{2} {\cal F}^{23\dot a}+12 \partial_{\dot b} {\cal F}^{\dot b3\dot a}=12 \partial_{\dot b} {\cal F}^{\dot b3\dot a}
\ee 
implies 
\be 
{\cal F}_{3\dot a\dot b}= \epsilon_{123\dot a\dot b\dot c}\partial^{\dot c}\Phi_{12}
\ee
Thus, use the gauge symmetry  $\delta A_{12}=\Phi_{12}$ we can completely remove the function $\Phi_{12}$ and find the self-dual relation.
\be {\cal F}_{3\dot a\dot b}=0
\ee 

Finally, through the  calculation  the field equation becomes
\be 0= {\delta S_{1+1+1+3}\over \delta A_{\dot a\dot b}}&=&-3(\partial_{\dot c} \tilde F^{\dot c\dot a\dot b}+\partial_{3} \tilde F^{3\dot a\dot b}+\partial_{2} \tilde F^{2\dot a\dot b}+\partial_{1} \tilde F^{1\dot a\dot b})\nn\\
&&+6\partial_{\dot c} {\cal F}^{\dot c\dot a\dot b}+6\partial_{3} {\cal F}^{3\dot a\dot b}+6\partial_{2} {\cal F}^{2\dot a\dot b}+3\partial_{1} \tilde F^{1\dot a\dot b}\nn\\
&=&6\partial_{\dot c} {\cal F}^{\dot c\dot a\dot b}+6\partial_{3} {\cal F}^{3\dot a\dot b}+6\partial_{2} {\cal F}^{2\dot a\dot b}+3\partial_{1} \tilde F^{1\dot a\dot b}=6 \partial_{\dot c} {\cal F}^{\dot c\dot a\dot b}
\ee
in which we have used the found self-duality rlations.   Above field equation has solution
\be{\cal F}_{\dot a\dot b\dot c}= \epsilon_{\dot a\dot b\dot c}\Phi(x^1, x^2, x^3)
\ee 
where $\Phi(x^1, x^2, x^3)$ is independent of the coordinates $x^{\dot a}$.  As  $\Phi(x^1, x^2, x^3)$ can be written as $\partial_i f^i(x^1, x^2, x^3)$, with $i=1,2,3$, the function $\Phi(x^1, x^2, x^3)$  can be absorbed by a field redefinition $A_{ij} \rightarrow A_{ij} + \epsilon_{ijk}f^k(x^1, x^2, x^3)$. Thus, we have another self-duality condition
\be {\cal F}_{\dot a\dot b \dot c}=0
\ee 
Together with other self-duality conditions we have found all self-duality conditions and complete the proof. 
%%%%%%%%%%%%%%%%%%%%
\subsubsection {Lorentz Invariance in D=1+1+1+3}
As covariant symmetry on 2D coordinates $x_{\dot a}$ is manifest we only need to examine transformations (I) mixing $x_1$ with $x_2$ and (II) mixing $x_1$ with $x_{\dot a}$.  Other  transformations needed to examine are just the exchange of $1\leftrightarrow 2$,  $1\leftrightarrow 3$ or $2\leftrightarrow 3$.

(I) For the mixing $x_1$ with $x_{\dot a}$ we shall consider the transformation
\be \delta x^{\dot a}&=&\omega^{\dot a 1}~x_1\equiv \Lambda^{\dot a}~x_1,\\
\delta x^{1}&=&\omega^{1\dot a }~x_{\dot a}=-\Lambda^{\dot a}~~x_{\dot a}=-\Lambda \cdot x
\ee
Define 
\be \Lambda \cdot L\equiv (\Lambda \cdot x)\partial_1 -x_1 (\Lambda \cdot \partial)
\ee
then we see that
\be
\delta  (F_{123}-\tilde F_{123})=(\Lambda \cdot L) (F_{123}-\tilde F_{123})+\Lambda^{\dot a} (F_{\dot a23}-\tilde F_{\dot a23})\\
\delta (F_{12\dot a}-\tilde F_{12\dot a})=(\Lambda \cdot L) (F_{12\dot a}-\tilde F_{12\dot a})+\Lambda^{\dot b} (F_{\dot b 2\dot a}-\tilde F_{\dot b 2\dot a})\\
\delta (F_{13\dot a}-\tilde F_{13\dot a})=(\Lambda \cdot L) (F_{13\dot a}-\tilde F_{13\dot a})+\Lambda^{\dot b} (F_{\dot b 3\dot a}-\tilde F_{\dot b 3\dot a})\\
\delta (F_{1\dot a\dot b}-\tilde F_{1\dot a\dot b})=(\Lambda \cdot L) (F_{1\dot a\dot b}-\tilde F_{1\dot a\dot b})+\Lambda^{\dot c} (F_{\dot c\dot a\dot b}-\tilde F_{\dot c\dot a\dot b})
\ee
which are zero for self-dual theory and the non-covariant action gives field equations with 6d Lorentz transformation mixing $x_1$ with $x_{a}$.
%%%%%%%%%%

(II) For the mixing $x_1$ with $x_2$ we shall consider the transformation
\be \delta x^{1}&=&\omega^{12}~x_2\equiv \Lambda~x_2,\\
\delta x^{2}&=&\omega^{21}~x_1=-\Lambda~x_1
\ee
Define 
\be \Lambda \cdot L\equiv \Lambda x_2 \partial_1 -x_2 \Lambda \partial_1
\ee
then we see that
\be
\delta  (F_{123}-\tilde F_{123})&=&(\Lambda \cdot L) (F_{123}-\tilde F_{123})\\
\delta (F_{12\dot a}-\tilde F_{12\dot a})&=&(\Lambda \cdot L) (F_{12\dot a}-\tilde F_{12\dot a})\\
\delta (F_{13\dot a}-\tilde F_{13\dot a})&=&(\Lambda \cdot L) (F_{13\dot a}-\tilde F_{13\dot a})-\Lambda(F_{23\dot a}-\tilde F_{23\dot a})\\
\delta (F_{1\dot a\dot b}-\tilde F_{1\dot a\dot b})&=&(\Lambda \cdot L) (F_{1\dot a\dot b}-\tilde F_{1\dot a\dot b})-\Lambda(F_{2\dot a\dot b}-\tilde F_{2\dot a\dot b})
\ee
which are zero for self-dual theory and the non-covariant action gives field equations with 6d Lorentz transformation mixing $x_1$ with $x_2$.
\\

In summary, we have found the non-covariant action of self-dual 2-form in decomposition $D=1+1+1+3$ and checked  that the non-covariant action gives field equations with 6d Lorentz transformation. 

%%%%%%%%%%%%%%%%%%%%
\subsection {Decomposition in Other Spacetime}
We have seen that the Lagrangian of self-dual 2-form gauge field could be expressed as many different formulations. The extension to other form is straightforward. \\

  For example, the self-dual 4-form gauge field in 10 D can be decomposed as $D=D_1+D_2$, which is proved in [14].  For the decompositions $D=1+9$, see table 4, the Lagrangian can be expressed as 
\be L_{1+9} = -{1\over4}  \sum L_{1\dot a\dot b\dot c\dot d}
\ee
with $\dot a=(2,\cdot \cdot \cdot ,10)$.
\\
\\
{Table 4: Lagrangian in decompositions: $D=1+9$ and $D=1+1+8$.}\\\\
\scalebox{1}{\hspace{3cm}\includegraphics{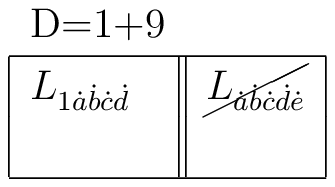}}\scalebox{1}{\hspace{2cm}\includegraphics{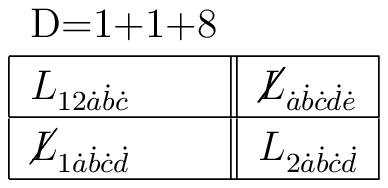}}
\\
\\
  The decomposition  in $D=D_1+D_2+D_3$  can also be performed  as before.  In the case of $D=1+1+8$ then, as that in section 3.1, the spacetime index $A$ is decomposed as $A=(1,2 ,\dot a)$, with $\dot a=(3,\cdot \cdot \cdot ,10)$, and  $L_{ABCEF}=(L_{12\dot a\dot b\dot c}, L_{\dot a \dot b\dot c\dot d\dot e}, L_{1\dot a\dot b\dot c\dot d}, L_{2\dot a\dot b\dot c\dot d})$.  From table 4 we see that Lagrangian can be expressed as 
\be L_{1+1+8} = -{1\over4} \Big(6\sum L_{12\dot a\dot b\dot c}+4 \sum L_{2\dot a\dot b\dot c\dot d}\Big)
\ee
Choosing $L_{12\dot a\dot b\dot c}+L_{2\dot a\dot b\dot c\dot d}$ is just the decomposition $D=1+9$ in [14].  The proof of self-dual relation is the same as those in section 3.1.   
\\

In conclusion, there are many different formulations of the self-dual gauge field.  As the decomposition has many kind it seems not easy to provide a general proof of the  self-dual relation in there and we have used some examples to illuminate the property.
%%%%%%%%%%%%%%%%%%
\section {Conclusion}  
In this paper we have first reviewed the Lagrangian of self-dual gauge theory in various non-covariant formulations. Then, we see a simple rule in there and use it to present some new  Lagrangian of  non-covariant forms of self-dual gauge theory. We see that the existence of  gauge symmetry $\delta A= \Phi$ in the Lagrangian play important role of the self-dual relation.  Using this as the guiding principle we have found many different formulations.  It is interesting to see that in some cases  it remains only one possible choice for the specified decomposition.   Especially, we have followed the prescription in [14]  to prove the self-dual property in the new Lagrangian in a detailed way. We also have followed the method of Perry and Schwarz in to show that these new non-covariant actions give field equations with 6d Lorentz invariance. \\

Finally, the covariant form in each decomposition may be found by following the method in [9,10,13] and we leave the study in future research.  It also remains to see whether the non-abelin self-dual gauge theory in 1+5 dimension  [15] could be decomposed in other way.
%%%%%%%%%%%%%%%%%%%
\\
\\
{\bf Acknowledgments} :  The author thanks Kuo-Wei Huang for interesting discussion in the initial stage of investigation. This work is supported in part by the Taiwan National Science Council. 
\\
%%%%%%%%%%%%%%%%%%%%%%%%%%%%
\\
\begin{center}{\Large \bf APPENDIX} \end{center}
\begin{appendix}
\section {Coordinate Transformation}
In this appendix we evaluate in detail the Lorentz transformation of 2-form field strength. 

Under the coordinate transformation :  $x_a\rightarrow x_a+\delta x_a$ we consider the  tensor  field $F_{\bar M\bar N\bar P}(x_a)$ defined by
\be
 F_{MNP}(x_a)&\rightarrow &F_{\bar M\bar N\bar P}(x_a+\delta x_a)\equiv{\partial x^Q\over \partial \bar x^{ M}}{\partial x^R\over \partial \bar x^{ N}}{\partial x^S\over \partial \bar x^{ P}}F_{QRS}(x_a+\delta x_a)\nn\\
&\approx&F_{MNP}(x_a+\delta x_a)+{\partial x^Q\over \partial \bar x^{ M}}{\partial x^R\over \partial \bar x^{ N}}{\partial x^S\over \partial \bar x^{ P}}F_{QRS}(x_a)\ee
For the transformation mixing between $x_1$ with $x_\mu$ ($\mu\neq1$) the relation $\delta x_a= \omega_{ab}x^b$ leads to $\delta x_1=-\Lambda_\mu x^\mu$ and  $\delta x_\mu=\Lambda_\mu x^1$ in which we define $\omega_{\mu 1}=-\omega_{1\mu}\equiv \Lambda_\mu$. 

The orbital part of transformation [10] is defined by 
\be \delta_{orb}F_{MNP}&\equiv&F_{MNP}(x_a+\omega_{ab}x^b)-F_{MNP}(x_a)\approx[\delta x_a] \cdot \partial^a F_{MNP}\nn\\
&=&[\omega_{ab}x^b \cdot \partial^a ] F_{MNP}= [\omega_{1\mu} x^\mu \partial^1]  F_{MNP}+[\omega_{\mu 1} x_1 \partial^\mu ] F_{MNP}\nn\\
&=&[\Lambda_\mu x^\mu \partial^1] F_{MNP}-x^1[\Lambda_\mu \partial^\mu ] F_{MNP} \nn\\
&=& [(\Lambda\cdot x)\partial^1-  x^1 (\Lambda \cdot\partial)] F_{MNP}\equiv  (\Lambda\cdot L) F_{MNP} 
\ee
Note that $\delta_{orb}$ is independent of index $MNP$ and is universal for all type tensor.  

The spin part of transformation [10]  becomes
\be \delta_{spin}H_{\mu\nu\lambda}&\equiv&{\partial x^Q\over \partial x^{\mu}}{\partial x^R\over \partial x^{\nu}}{\partial x^S\over \partial x^{\lambda}}H_{QRS}(x)- H_{\mu\nu\lambda}\nn\\
&\approx&\Big[{\partial (\delta x^1)\over \partial x^{\mu}}{\partial x^R\over \partial x^{\nu}}{\partial  x^S\over \partial x^{\lambda}}H_{1 RS}(x)+{\partial (\delta x^{\sigma})\over \partial x^{\mu}}{\partial x^R\over \partial x^{\nu}}{\partial  x^S\over \partial x^{\lambda}}H_{\sigma RS}(x)\Big]+\Big[{\partial x^Q\over \partial x^{\mu}}{(\delta x^1)\over \partial x^{\nu}}{\partial  x^S\over \partial x^{\lambda}}H_{Q1 S}(x)
\nn\\
&&+{\partial x^Q\over \partial x^{\mu}}{(\delta x^{\sigma})\over \partial x^{\nu}}{\partial  x^S\over \partial x^{\lambda}}H_{Q\sigma S}(x)\Big]+\Big[{\partial x^Q\over \partial x^{\mu}}{\partial x^R\over \partial x^{\nu}}{\partial  (\delta x^1)\over \partial x^{\lambda}}H_{QR1}(x)
+{\partial x^Q\over \partial x^{\mu}}{\partial x^R\over \partial x^{\nu}}{\partial  (\delta x^{\sigma})\over \partial x^{\lambda}}H_{QR\sigma}(x)\Big]
\nn\\
&=&{\partial (\delta x^1)\over \partial x^{\mu}}{\partial x^R\over \partial x^{\nu}}{\partial  x^S\over \partial x^{\lambda}}H_{1 RS}(x)+{\partial x^Q\over \partial x^{\mu}}{(\delta x^1)\over \partial x^{\nu}}{\partial  x^S\over \partial x^{\lambda}}H_{Q1 S}(x)+{\partial x^Q\over \partial x^{\mu}}{\partial x^R\over \partial x^{\nu}}{\partial  (\delta x^1)\over \partial x^{\lambda}}H_{QR1}(x)
\nn\\
&=&\Big[- \Lambda_\mu H_{1\nu \lambda}\Big]+\Big[-\Lambda_\nu  H_{\mu1\lambda}\Big]+\Big[-\Lambda_\lambda H_{\mu\nu1}\Big]
\ee
and $\delta H_{\mu\nu\lambda}=\delta_{orb}H_{\mu\nu\lambda}+\delta_{spin}H_{\mu\nu\lambda}$

In a same way 
\be \delta_{spin}H_{\mu\nu1}&\equiv&{\partial x^Q\over \partial x^{\mu}}{\partial x^R\over \partial x^{\nu}}{\partial x^S\over \partial x^{1}}H_{QRS}(x)- H_{\mu\nu1}\nn\\
&\approx&\Big[{\partial (\delta x^1)\over \partial x^{\mu}}{\partial x^R\over \partial x^{\nu}}{\partial  x^S\over \partial x^{1}}H_{1 RS}(x)+{\partial (\delta x^{\sigma})\over \partial x^{\mu}}{\partial x^R\over \partial x^{\nu}}{\partial  x^S\over \partial x^{1}}H_{\sigma RS}(x)\Big] +\Big[{\partial x^Q\over \partial x^{\mu}}{(\delta x^1)\over \partial x^{\nu}}{\partial  x^S\over \partial x^{1}}H_{Q1 S}(x)\nn\\
&&{\partial x^Q\over \partial x^{\mu}}{(\delta x^{\sigma})\over \partial x^{\nu}}{\partial  x^S\over \partial x^{1}}H_{Q\sigma  S}(x)\Big]+\Big[{\partial x^Q\over \partial x^{\mu}}{\partial x^R\over \partial x^{\nu}}{\partial  (\delta x^1)\over \partial x^{1}}H_{QR1}(x)+{\partial x^Q\over \partial x^{\mu}}{\partial x^R\over \partial x^{\nu}}{\partial  (\delta x^\lambda)\over \partial x^{1}}H_{QR\lambda}(x)\Big]
\nn\\
&=&{\partial (\delta x^1)\over \partial x^{\mu}}{\partial x^R\over \partial x^{\nu}}{\partial  x^S\over \partial x^{1}}H_{1 RS}(x)+{\partial x^Q\over \partial x^{\mu}}{(\delta x^1)\over \partial x^{\nu}}{\partial  x^S\over \partial x^{1}}H_{Q1 S}(x)+{\partial x^Q\over \partial x^{\mu}}{\partial x^R\over \partial x^{\nu}}{\partial  (\delta x^\lambda)\over \partial x^{1}}H_{QR\lambda}(x)
\nn\\
&=&0+0+\Lambda^\lambda H_{\mu\nu\lambda}
\ee
and $\delta H_{\mu\nu1}=\delta_{orb}H_{\mu\nu1}+\delta_{spin}H_{\mu\nu1}=(\Lambda\cdot L)H_{\mu\nu1}+\Lambda^\lambda H_{\mu\nu\lambda}$.
\end{appendix}
%%%%%%%%%%%%%%%%%%%%%%
\\
\begin{center} {\bf REFERENCES}\end{center}
%%%%%%%%%%%%%%%%%%%%%%
\begin{enumerate}
\item M. Green, J. Schwarz and E. Witten,``Superstring theory'', Cambridge University Press, Cambridge, 1987.
\item C.G. Callan, J.A. Harvey, and A. Strominger, Nucl. Phys. B367, 60 (1991); E. Witten,`` Five brane effective action'', hep-th/9610234 [hep-th/9610234].
\item L. Alvarez-Gaum and E. Witten, Nuc. Phys. B234 (1983) 269.
\item N. Marcus and J.H. Schwarz, Phys. Lett. 115B (1982) 111;\\J. H. Schwarz and A. Sen, ``Duality symmetric actions," Nucl. Phys. B 411, 35 (1994) [arXiv:hep-th/9304154].
\item R. Floreanini and R. Jackiw,``Selfdual fields as charge density solitons,'' Phys. Rev. Lett. 59 (1987) 1873.
\item M. Henneaux and C. Teitelboim, ``Dynamics of chiral (selfdual) p forms,''  Phys. Lett. B 206 (1988) 650.
\item W. Siegel,``Manifest Lorentz invariance sometimes requires nonlinearity,'' Nucl. Phys. B238 (1984) 307.
\item B. McClain, Y. S. Wu, F. Yu, Nucl. Phys. B343 (1990) 689; \\C. Wotzasek, Phys. Rev. Lett. 66 (1991) 129; \\F. P. Devecchi and M. Henneaux, Phys. Rev. D45 (1996)1606.
\item P. Pasti, D. Sorokin and M. Tonin, ``Note on manifest Lorentz and general coordinate invariance in duality symmetric models,'' Phys. Lett. B 352 (1995) 59
[arXiv:hep-th/9503182];\\ P. Pasti, D. Sorokin and M. Tonin, ``Duality symmetric actions with manifest space-time symmetries,'' Phys. Rev. D 52 (1995) R4277 [arXiv:hep-th/9506109]; \\P. Pasti, D. Sorokin and M. Tonin, ``Space-time symmetries in duality symmetric models,'' [arXiv:hep-th/9509052]; P. Pasti, D. P. Sorokin, M. Tonin, ``On Lorentz invariant actions for chiral p-forms,'' Phys. Rev. D 55 (1997) 6292 [arXiv:hep-th/9611100].
\item M. Perry and J. H. Schwarz, ``Interacting Chiral Gauge Fields in Six Dimensions and Born-Infeld Theory" Nucl.Phys. B489 (1997) 47-64 [arXiv:hep-th/9611065]
\item P. M. Ho, Y. Matsuo, ``M5 from M2'', JHEP 0806 (2008) 105 [arXiv: 0804.3629 [hep-th]];  \\P.M. Ho, Y. Imamura, Y. Matsuo, S. Shiba, ``M5-brane in three-form flux and multiple M2-branes,'' JHEP 0808 (2008) 014, [arXiv:0805.2898 [hep-th]].
\item  J. Bagger and N. Lambert, ``Modeling multiple M2'' Phys. Rev. D 75 (2007) 045020 [arXiv:hep-th/0611108];\\
J.A. Bagger and N. Lambert, ``Gauge Symmetry and Supersymmetry of Multiple M2-Branes,'' Phys. Rev. D 77 (2008) 065008 arXiv:0711.0955 [hep-th];\\
J. Bagger and N. Lambert, ``Comments On Multiple M2-branes,'' JHEP 0802 (2008) 105, arXiv:0712.3738 [hep-th]; \\A. Gustavsson, ''Algebraic structures on parallel M2-branes,'' Nucl. Phys. B 811 (2009) 66, arXiv:0709.1260 [hep-th].
\item Pasti, Samsonov, Sorokin and Tonin,``BLG-motivated Lagrangian formulation for the chiral two-form gauge field in D = 6 and M5-branes'' Phys. Rev. D 80 (2009) 086008 [arXiv:0907.4596 [hep-th]].
\item W.-M. Chen and P.-M. Ho,``Lagrangian Formulations of Self-dual Gauge Theories in Diverse Dimensions'' Nucl. Phys. B837 (2010) 1 [arXiv:1001.3608 [hep-th]]. 
 \item P.-M. Ho, K.-W. Huang and Y. Matsuo,`` A Non-Abelian Self-Dual Gauge Theory in 5+1 Dimensions'' JHEP 1107 (2011) 021 [arXiv:1104.4040  [hep-th]].   
\end{enumerate}
\end{document}